\documentclass[twocolumn,showpacs,floatfix,superscriptaddress]{revtex4-1}
\usepackage{graphicx,epsfig,dcolumn,amssymb,amsmath}
\usepackage{booktabs}

\usepackage{color}

\begin{document}

\title{Computing Optical Properties of Ultra-thin Crystals}

\author{H. Sahin}
\email{hasan.sahin@uantwerpen.be}
\affiliation{Department of Physics, University of Antwerp, Groenenborgerlaan 
171, 2020 Antwerp, Belgium}

\author{E. Torun}
\affiliation{Department of Physics, University of Antwerp, Groenenborgerlaan 
171, 2020 Antwerp, Belgium}

\author{C. Bacaksiz}
\affiliation{Department of Physics, Izmir Institute of Technology, 35430 Izmir, Turkey}

\author{S. Horzum}
\affiliation{Department of Physics, University of Antwerp, Groenenborgerlaan 171, 2020 Antwerp, 
Belgium}
\affiliation{Department of Engineering Physics, Faculty of Engineering, Ankara University, 06100 
Ankara, Turkey}

\author{J. Kang}
\affiliation{Department of Physics, University of Antwerp, Groenenborgerlaan 171, 2020 Antwerp, 
Belgium}

\author{R. T. Senger}
\affiliation{Department of Physics, Izmir Institute of Technology, 35430 Izmir, Turkey}

\author{F. M. Peeters}
\affiliation{Department of Physics, University of Antwerp, Groenenborgerlaan 171, 2020 Antwerp, 
Belgium}
\date{\today}
\date{\today}


\begin{abstract}
An overview is given of recent advances in experimental and theoretical understanding of optical 
properties of ultra-thin crystal structures (graphene, phosphorene, silicene, MoS$_{2}$, 
MoSe$_{2}$, WS$_{2}$, WSe$_{2}$, h-AlN, h-BN, fluorographene, graphane). Ultra-thin crystals are 
atomically-thick layered crystals that have unique properties which differ from their 3D 
counterpart. Because of the difficulties in the synthesis of few-atom-thick  crystal structures, 
which are thought to be the main building blocks of future nanotechnology, reliable theoretical 
predictions of their electronic, vibrational and optical properties are of great importance. Recent 
studies revealed the reliable predictive power of existing theoretical approaches based on density 
functional theory (DFT). \end{abstract}

\maketitle

\section{Introduction}

Over the past decade, advances in materials science have shown 
that materials can have remarkably different properties when its dimensions are
reduced. In particular, the successful isolation of graphene in 
2004\cite{graphene} uncovered a new class of materials; ``\textit{two 
dimensional (2D) atomically-thin crystal structures}".\cite{heterovdw} 
Following the rapid development of both top-down and bottom-up synthesis 
approaches in recent years, more and more members have joined the family of 
ultra-thin 2D crystals, such as graphene derivatives (graphane, grapheneoxide, 
fluorographene and chlorographene), hexagonal BN and AlN, silicene, 
transition-metal chalcogenides (with formula MX$_{2}$ where M=Mo, W, Re and 
X=S, Se, Te), black phosphorus and metal hydroxides (Mg(OH)$_{2}$ and 
Ca(OH)$_{2}$). Many groups reported that in addition to their structural 
stability, 2D crystals also show a variety of exceptional electronic and 
optical properties. Despite the large amount of work that has been done on the 
structural and electronic properties of 2D materials, theoretical and 
experimental research on their optical properties are in their early stage. In 
this overview, following a brief introduction to experimental efforts on 2D 
crystals, we focus on computational methods and the most recent theoretical 
findings of their optical properties.

\begin{figure*}[htbp]
\includegraphics[width=15cm]{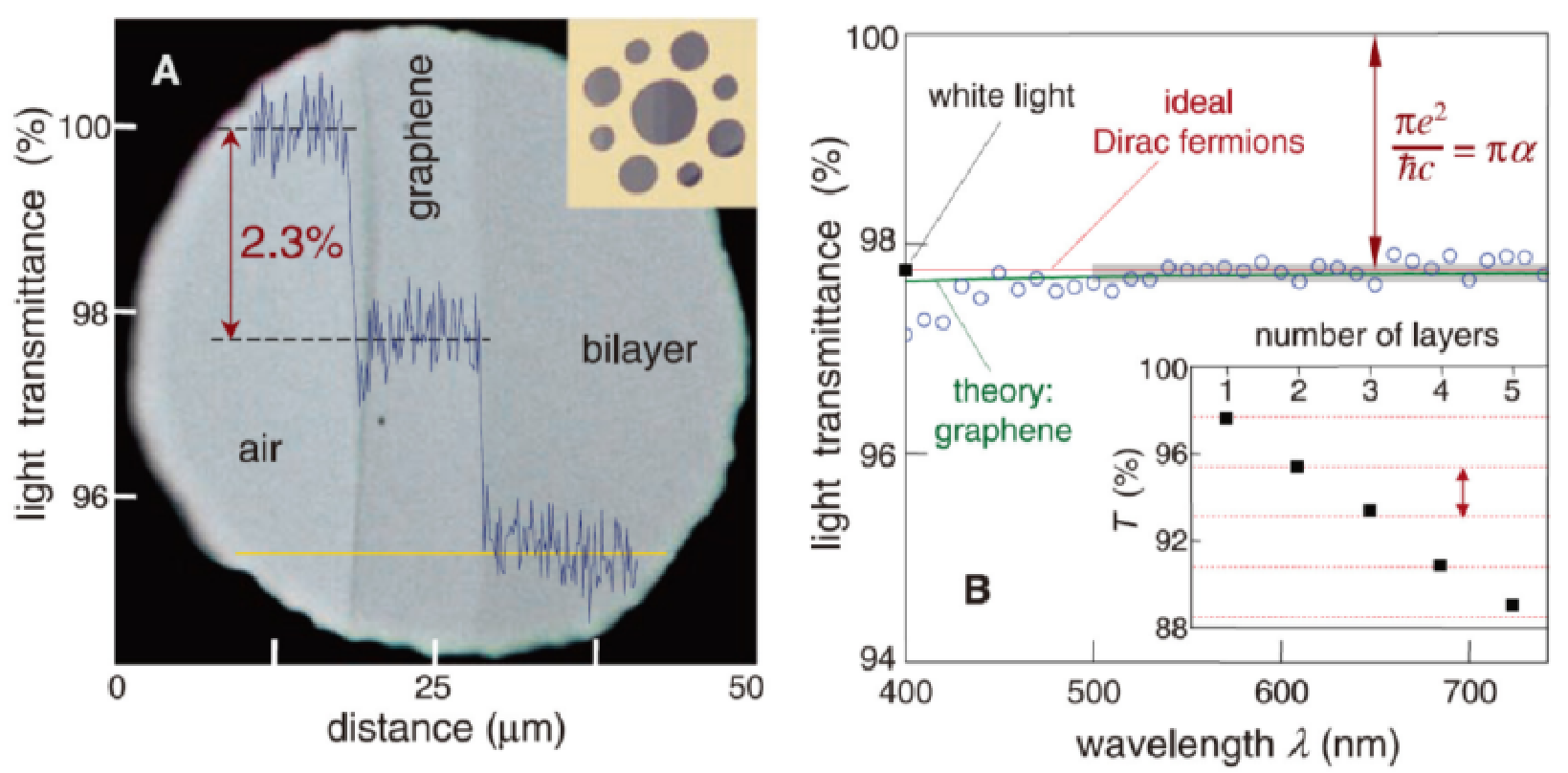}
\caption{\label{grapnene-absorption}
(a) Photograph of a 50-$\mu m$ aperture partially covered by graphene and its
bilayer. (b) Transmittance spectrum of single-layer graphene (open circles) and 
theoretically predicted results for two-dimensional Dirac fermions. 
Inset shows the transmittance of white light as a function of the
number of graphene layers (squares) [from Ref. 3].}
\end{figure*}

Experimentally, one of the first successful determinations of the optical 
reflectivity and transmission of graphene was performed by Nair \emph{\textit{et 
al}.}\cite{Nair} As shown in Figures \ref{grapnene-absorption}(a) and (b), the 
reflection of single layer graphene is very low, less than 0.1 \%  in the 
visible region. However, its optical absorption  $A$= $\pi\alpha$=2.3 \% is 
considerable, which is practically constant over the visible spectrum, where 
$\alpha$ is the fine structure constant 1/137. In addition, in multi-layer 
graphene, the optical absorption is proportional to the number of layers, i.e., 
each graphene layer contributes 2.3\% and the overall absorption is the 
superposition of those monolayers. However, such simple result may break down in 
the non-visible regions. In the work of Mak et al.\cite{mak08} 
two important aspects in the optical spectrum of graphene are: (i) possible 
many-body effects leading to a reduction of the optical conductivity at lower 
photon energies, and (ii) deviations from the linear 
dispersion that leads to an increase of the optical absorption at high energies.

Not only the absorption properties but also other topics related to its optical  
properties of graphene are under intensive investigations. Sun \textit{et  
al}.\cite{sun08} reported the first ultrafast spectroscopy of hot Dirac 
quasiparticles in epitaxial graphene in the region near the Fermi level. In the 
same year, infrared spectroscopy measurements  performed by Wang \textit{et 
al}.\cite{wang08} revealed that monolayers and bilayers of  graphene have strong 
interband transitions and that their optical transitions can be substantially 
modified through electrical gating, much like electrical transport in 
field-effect transistors. This gate dependence of the interband transitions  is 
an interesting tuning parameter for optically probing of the band structure of 
graphene. For monolayer graphene, it yields directly the linear band 
dispersion of Dirac fermions, whereas in bilayer graphene, a dominating van 
Hove singularity arising from interlayer coupling is observed. The strong and 
layer-dependent optical transitions of graphene and the tunability by simple 
electrical gating holds promise for new applications in infrared optics and 
optoelectronics. In the work of Xia \textit{et al}.\cite{xia-nnano} ultrafast 
photocurrent response measurements using a 1.55 $\mu$m excitation laser revealed 
a light intensity modulation with frequencies up to 40 GHz in single and 
few-layers graphene field-effect-transistors (FETs). They also concluded that the 
intrinsic bandwidth of such a graphene FET may lead to graphene based 
photodetectors that exceed 500 GHz.

Although graphene is a semi-metal in the monolayer form, it can exhibit 
photoluminescence (PL) by inducing a bandgap. Basically there are two 
approaches to open a bandgap in graphene. One is cutting graphene into 
nanoribbons and quantum dots, so that quantum confinement results in a gap 
opening. Pan \textit{et al}. prepared water-soluble graphene quantum 
dots that exhibit bright blue PL by hydrothermal (chemical) cutting of oxidized 
graphene sheets.\cite{pan-am} Jin \textit{et al}. further showed that the PL 
emission of graphene dots can be tuned through the charge transfer effect of 
functional groups.\cite{jin-acsnano} The other way to open a gap in graphene is 
by chemical or physical treatments which can reduce the connectivity of the 
$\pi$-electron network. For instance, by using an oxygen plasma treatment, 
Gokus \textit{et al}. observed strong PL in single-layer graphene.\cite{gokus} 
Moreover, Eda \textit{et al}. reported blue PL from chemically derived
grapheneoxide.\cite{eda} Chien \textit{et al}. demonstrated that it was 
possible to tune the PL in grapheneoxide suspensions from red to blue emission 
through the reduction of the surface oxide groups\cite{chien}, which could thus 
gradually change the ratio between  sp$^2$- and sp$^3$-bonded carbon atoms .

\begin{figure}
\includegraphics[width=8.0cm]{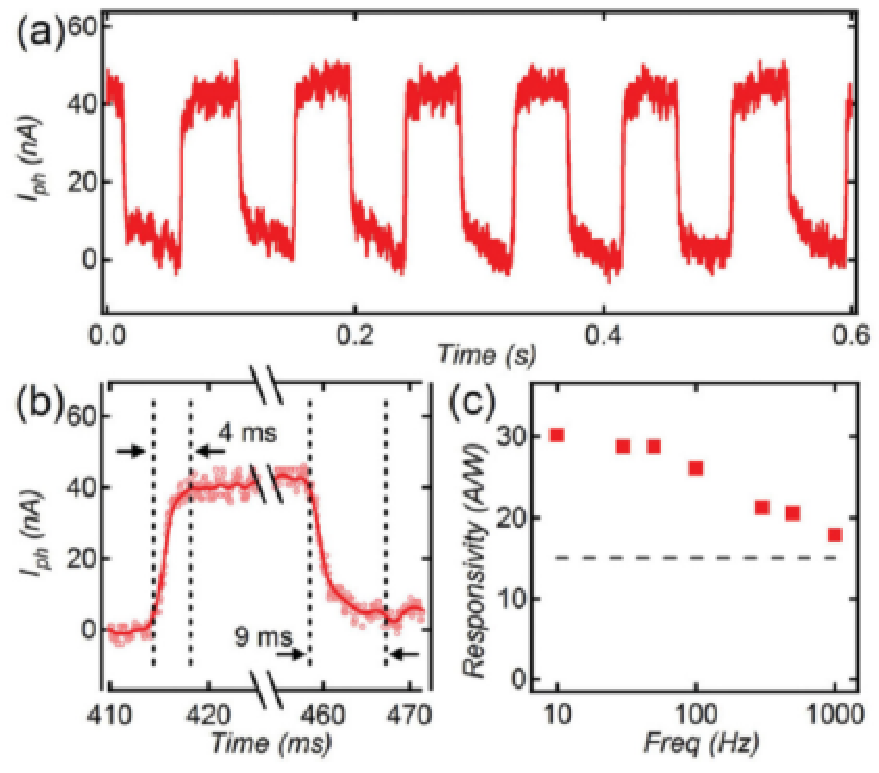}
\caption{\label{tis3} (a) Current response of TiS$_3$ field effect transistor 
under a 10 Hz mechanically modulated optical excitation. (b) Zoom on a single 
switching cycle at 10 Hz frequency. (c) Responsivity versus increasing 
modulation frequency for an excitation wavelength of 640 nm at 500 $\mu$W 
[from Ref. 14].}
\end{figure}

The optical properties of other ultra-thin crystals than graphene has also 
attracted a lot of attention. The in-plane dielectric function of four
different TMCs: MoS$_{2}$, MoSe$_{2}$, WS$_{2}$ and WSe$_{2}$ were determined by 
Li \textit{et al}.\cite{li14} through reflectance measurements at room 
temperature. They concluded that for all four TMC monolayers, the two lowest 
energy peaks in the reflectance spectrum correspond to excitonic features that 
stem from interband transitions at the K (K') point, while at higher photon 
energies the spectral broad response originates from higher-lying interband 
transitions. Island \textit{et al}.\cite{tis3-1} reported an ultrahigh 
photoresponse up to 2910 A W$^{-1}$ and fast switching times of about 4 ms with 
a cutoff frequency of 1000 Hz in a TiS$_3$ field effect transistor (see 
Figure \ref{tis3}). TiS$_3$ is also proven to be suitable for 
hydrogen photogeneration under visible light.\cite{tis3-2}

For TMC semiconductors, measurement of PL is one of the most efficient ways to 
find out their characteristics. In early experimental studies on TMCs, Mak 
\textit{et al}.\cite{mak-mos2} and Splendiani et al.\cite{splendiani} showed 
that bulk MoS$_{2}$ which exhibits PL, displays a strong peak intensity 
increase when the crystal is thinned down to a monolayer. The increasing PL 
intensity was explained as a consequence of an indirect to direct bandgap 
transition when going from multi to monolayer graphene. Subsequently similar  
indirect-to-direct bandgap crossover has been reported for other TMCs such as 
MoSe$_{2}$\cite{Tonndorf,sef-1}, WSe$_{2}$\cite{Tonndorf} and 
WS$_{2}$\cite{ws2-pl}.

In conventional semiconductors, increasing temperature or introducing defects 
usually reduces the PL strength. Interestingly, studies by Tongay \textit{et 
al}. showed the opposite behavior in TMCs.\cite{sef-1,sef-2} They found that 
the PL of few-layer MoSe$_2$ was much enhanced with increasing temperature, due 
to the thermally driven crossover from indirect towards a 
direct bandgap.\cite{sef-1} In another work, they observed an 
increase in the overall PL intensity as a result of the generation of vacancies in 
monolayer TMCs when put in a gas environment (such as N$_{2}$).\cite{sef-2} The 
interaction between the gas molecules and the defect sites, which caused a 
transition of exciton population from charged to neutral excitons, was 
responsible for the PL enhancement. In monolayer TMCs, due to the reduced 
screening of the Coulomb interaction, the exciton binding energy is much larger 
than in bulk materials. Ugeda \textit{et al}. reported an exciton binding 
energy of 0.55 eV for monolayer MoSe$_2$ by means of scanning tunneling 
spectroscopy and the two-particle exciton transition energy using PL 
spectroscopy.\cite{ugeda}

In addition, many honeycomb lattice TMCs such as MoS$_2$ and WSe$_2$ showed 
rich valley-polarized optical properties. The inequivalent valleys in the 
Brillouin zone could be selectively addressed using circularly-polarized light 
fields.\cite{cao,zeng,jones} Jones \textit{et al}.\cite{jones} reported 
the experimental generation and detection of valley coherence in WSe$_{2}$. 
Their study was based on the fact that excitons localized in a single valley emit 
circularly polarized photons. Linear polarization can only be generated 
through recombination of an exciton which is in a coherent superposition of the two 
valley states. Furthermore, Srivastava \textit{et al}. realized a valley Zeeman 
effect in WSe$_2$.\cite{Srivastava} They showed strong anisotropic lifting of 
the degeneracy of the valley pseudospin degree of freedom by using an external 
magnetic field.

\begin{figure}[htbp]
\includegraphics[width=6cm]{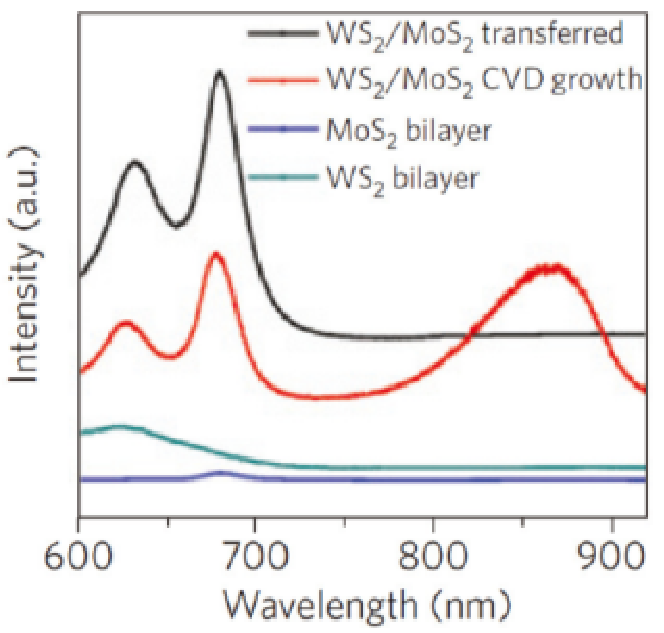}
\caption{\label{gong14} PL spectrum of a CVD-grown WS$_2$/MoS$_2$ bilayer made 
by mechanical transfer, and CVD-grown MoS$_2$ and WS$_2$ bilayers [from Ref. 28].}
\end{figure}

Most recent experimental efforts have also revealed that synthesis 
of heterostructures made of monolayer crystal structures can be achieved such 
as van der Waals heterostructures made of 
TMC monolayers which exhibit novel optical properties. Tongay \textit{et al}. 
reported the realization of large-area (tens of micrometers) heterostructures 
of CVD-grown WS$_2$ and MoS$_2$ monolayers.\cite{sef-mos2-ws2} They showed that 
the interlayer coupling can be tuned by vacuum thermal annealing. As the 
coupling strength increases, the photoluminescence gradually changes from a 
superposition of the spectra of MoS$_2$ and WS$_2$ monolayers to the coupled 
spectrum, in which charged exciton (also called trions) recombination and 
phonon-assisted indirect-bandgap transition dominate the light emission. Gong 
\textit{et al}. also observed interlayer exciton transition in a
MoS$_2$/WS$_2$ heterostructure, which leads to an additional PL peak at 1.4 
eV which is shown in Figure \ref{gong14}.\cite{gong} Similar interlayer exciton 
transition in MoSe$_2$/WSe$_2$ heterostructure was also reported by Rivera 
\textit{et al}., and the interlayer exciton had a long lifetime of 1.8 ns, an 
order of magnitude longer than intralayer excitons.\cite{Rivera} Besides their PL 
properties, Huo \textit{et al}. demonstrated enhanced optoelectronic 
performances of multilayer MoS$_2$/WS$_2$ heterostructure transistors, whose 
photoresponsivity is three orders of magnitude larger than for devices made of 
monolayer MoS$_2$.\cite{Huo}  Several studies have reported the observation of 
 a photovoltaic response in MoS$_2$/WSe$_2$ 
heterostructures. \cite{rcheng,chlee,furchi} The collection of the 
photoexcited carriers can be enhanced by sandwiching MoS$_2$/WSe$_2$ between 
graphene\cite{chlee}. Furthermore, Withers \textit{et al}. demonstrated a new 
type of light-emitting diode based on van der Waals heterostructures, which 
consist of graphene, hexagonal boron nitride and various semiconducting TMCs 
such as MoS$_2$.\cite{Withers} The devices showed a quantum efficiency up to 8.4 
\%, and the emission spectrum can be tuned by choosing different semiconducting 
layers.

\begin{figure*}[htbp]
\includegraphics[width=11cm]{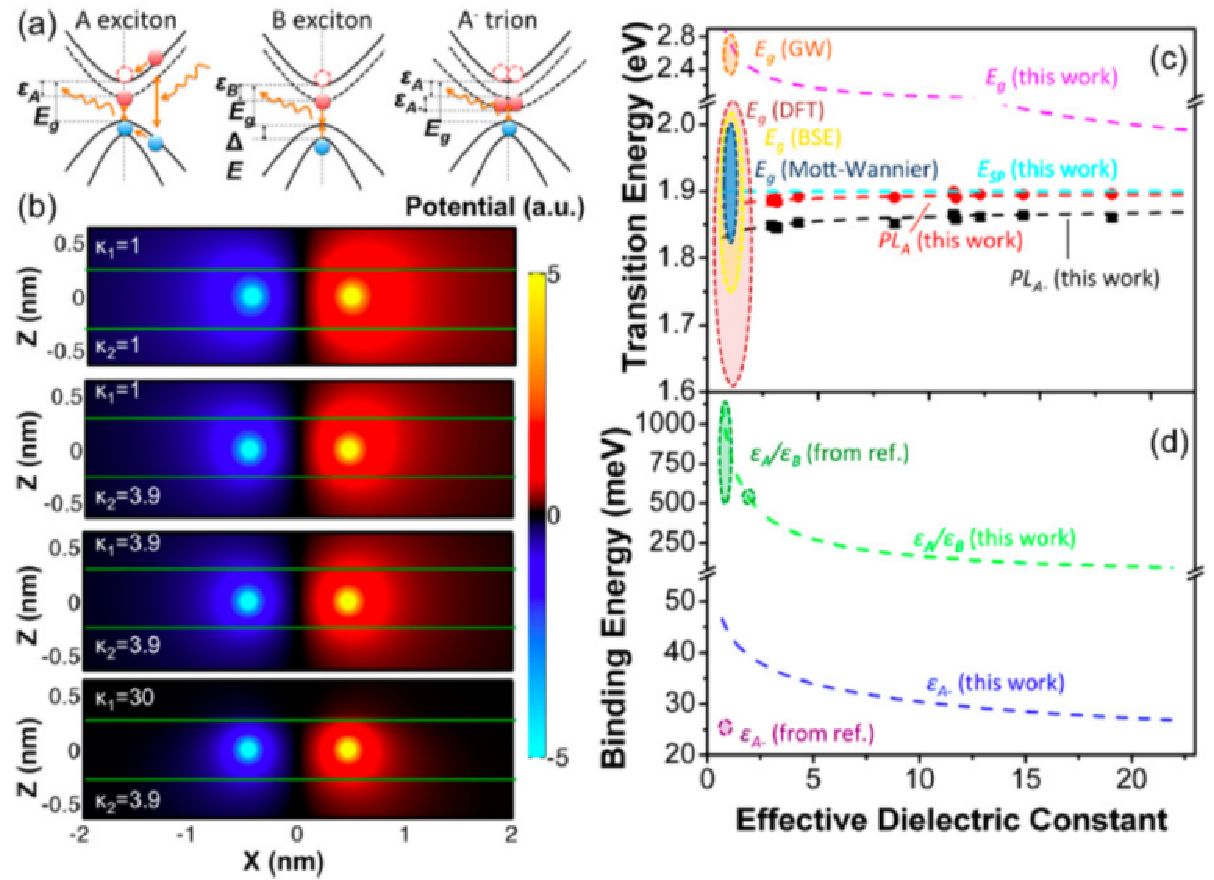}
\caption{\label{new} (a) The radiative transitions related to excitons at the 
$K$-point in the BZ. E$_g$, $\Delta$, $\epsilon_{A}$, 
$\epsilon_{B}$,$\epsilon_{A^{-}}$ correspond to band gap, valence band 
splitting, binding energy of A, B, exciton and A$^{-}$ trion, respectively. (b) 
Contour plots of the Coulomb potential in the middle of the layer for four 
different dielectric configurations. (c) The PL peak energies of A exciton 
(red) and A$^{-}$ trion (black) as a function of the effective dielectric 
constant. (d) The binding energies of A, B exciton (green) and A$^{-}$ trion 
(blue) as function of the effective dielectric constant [from Ref. 44].}    
\end{figure*}

Furthermore, excitonic properties of 2D crystal structures 
show significant deviations from their bulk forms. While the exciton binding 
energy varies between few to 100 meV in an usual semiconductor, 2D TMDs have 
exciton binding energies of several hundred meV.\cite{Chernikov,Ye,He,ugeda} 
Such a strong enhancement in the excitonic 
binding energy is attributed to two main effects: (i) the confinement forces 
that localize the electron and the hole at the same place (which causes  
a stronger binding energy), and (ii) the dielectric 
screening of the Coulomb interaction is significantly reduced due to the 
dielectric mismatch effect as compared to the bulk. Additionally, it 
was demonstrated that absorption of a photon having energy larger than the 
bandgap in a 2D semiconductor can also led to the formation of biexcitons and 
trions (negatively charged electron-exciton or positively charged 
hole-exciton).\cite{jones,trion-mak,Ross,Mitioglu,Mai,Shang,You} As 
experimentally shown by Lin \textit{et al.}\cite{lin}  (see 
Figure \ref{new}(b)) for 2D materials that have high surface-to-volume ratio 
the distribution of the Coulomb interaction and therefore the behavior of 
excitons are strongly modified by the dielectric environment. It was observed 
that different dielectric environments can tune the binding energies of 
excitons and trions by one order of magnitude. The relative binding energy of 
exciton and trion of the single-layer MoS$_2$ are larger when the system is in 
vacuum and they decrease with the effective dielectric constant as shown in 
Figure \ref{new}(d). In another recent study the screening properties of 
two-dimensional semiconductors and layered structures were investigated by 
Latini \textit{et al.} It was shown that while the screened interaction in the 
generalized Mott-Wannier model yields results almost identical to those of the 
strict 2D model (\textit{ab-initio} calculations) for exciton binding 
energies, it may fail for heterostructures and when supported 2D materials are 
present. They also showed that a quasi-2D model, that takes into account the 
finite extension of the 2D material in the out-of-plane direction, provides a 
much broader applicability. \cite{latini}

Following the above-mentioned experimental works on the characterization of the
optical properties of ultra-thin crystals, we will focus here on the 
theoretical efforts and recent findings on the optical properties of two 
dimensional atomic-thin materials. Next section includes a detailed overview of 
the optical properties of: (i) elemental monolayers of group IV and V, (ii) 
derivatives of graphene, (iii) transition-metal chalcogenides, (iv) III-V 
binary compounds, and (v) heterostructures.

\section{Theoretical Determination of Optical properties of Ultra-thin Crystals}

Despite the success of LDA\cite{LDA} and GGA\cite{GGA1,GGA2} approximations in describing the 
exchange and correlation 
interactions and resulting electronic band dispersion of structures, electronic and optical energy 
bandgap values are mostly 
underestimated by DFT. Since simple Kohn-Sham eigenvalues give inaccurate quasiparticle energy 
spectrum for the structure,  
Kohn-Sham DFT can not explain properly the phenomena which include electronic excitations like 
photoemission or 
photoabsorption since the Kohn-Sham.\cite{onida} However, quite reliable results can be obtained by 
using beyond-DFT 
approaches such as HSE, GW and BSE.

 One of the deficiencies of DFT arises from local exchange functionals, that give smaller 
ionization potential and electron 
affinity values. To solve this systematical error one can use hybrid functionals such as HSE; the 
short range part of the 
exchange term is defined as a mixture of exact short-range exchange from the Fock theory and  
semilocal long-range exchange 
of GGA. It was shown that nonlocal potetntials generated by HSE functionals accurately describe the 
band gap of 
semiconductos\cite{HSE}.

Although the GGA+HSE and LDA+HSE methods are successful in finding electronic bandgap, inclusion of 
the self energy of the 
electron ($i$GW)  is essential for cases in which the screening effects are large. This can be done 
by using the GW 
approximation\cite{GW1,GW2,GW3,GW4} where G and W stand for the single particle Green's function 
and the screened Coulomb 
interaction, respectively. Briefly, the self energy of a particle is the energy which originates 
from the disturbance on the 
environment by the particle itself, and the particle with its disturbance is the quasiparticle 
(QP). There are several 
methods to calculate the QP energies, such as the G$_{0}$W$_{0}$ (single shot for the G and W), 
GW$_{0}$ (several iterations 
over the G and no iterations over the W), GW (several iterations over the G and W) as well as the 
scGW (fully 
self-consistent). These methods are implemented in softwares ABINIT,\cite{ABINIT} 
Yambo,\cite{Yambo} 
QuantumEspresso,\cite{Q.E.}  and VASP.\cite{vasp1,vasp2,vasp3,vasp4}

The inclusion of screening effects is sufficient to obtain the accurate electronic structure of a 
material. However, to 
obtain accurate optical spectrum, such as absorption and photoluminescence, the excitonic states 
must be taken into account. 
Because, in the optical excitation mechanism, an electron usually makes a bound state with a hole 
rather than moving 
back-and-forth between the QP states. Furthermore, considering only the excitations between the QP 
states underestimates the 
strength of the low energy excitations and overestimates the high energy excitations. Therefore, 
the Bethe-Salpeter equation 
(BSE)\cite{BSE1,BSE2} must be solved for the two-particle Green's function of electron-hole pairs. 
The BSE is expressed in 
terms of the wave functions and energies that are obtained from the QP calculation. The commonly 
used program packages which 
are mentioned above are also capable to solve the BSE.

\subsection{Elemental Monolayers of Group IV and V}

\begin{figure}[htbp]
\includegraphics[width=7.5cm]{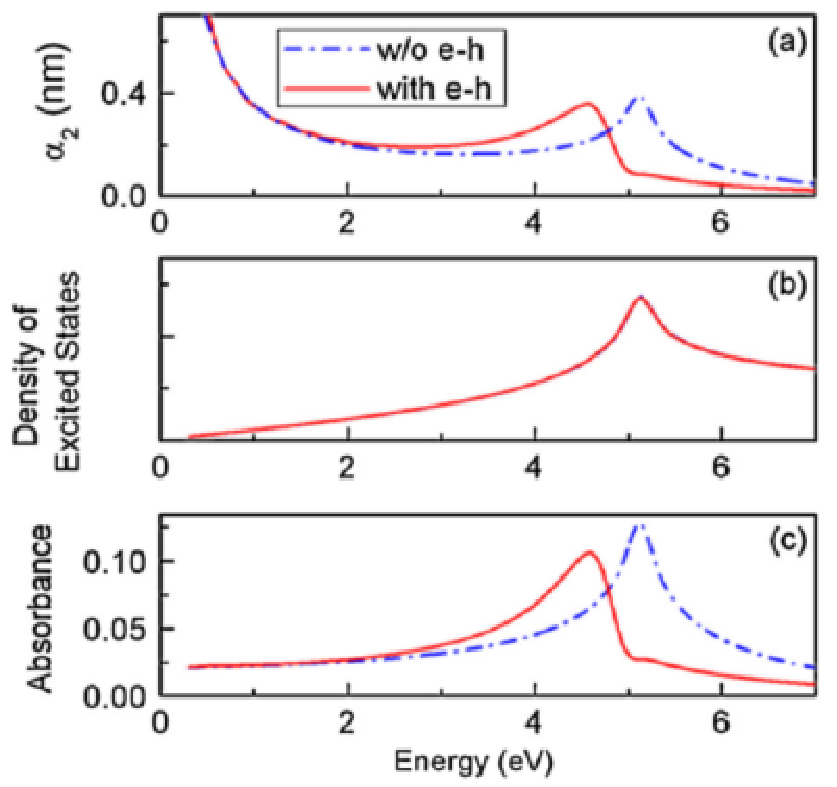}
\caption{\label{yang_prl_2009}
(a) Optical absorption spectrum, (b) density of excited states, (c) absorbance 
of graphene, with and without excitonic effects included [from Ref. 64].}
\end{figure}

The pioneer of the 2D materials is graphene which is an elemental monolayer 
structure of carbon atoms. Beside its unusual mechanical and electronic 
properties, graphene displays many interesting optical features, such as a 
constant optical conductivity and tunable optical absorbance. After the 
synthesis of graphene in 2004, the first theoretical attempt to understand 
graphene's optical properties within the first-principles method was made by 
Yang \textit{\textit{et al}.} in 2009. They investigated the optical properties 
of single and bilayer graphene including many-electron effects.\cite{Yang} It 
is commonly believed that excitonic effects are not important for metals 
because of the high screening. However, they reported that the excitonic 
effects significantly change the absorption spectrum (where the main peak 
position is red-shifted) in the energy regime near the van Hove singularity 
shown in Figure \ref{yang_prl_2009}, whereas excitonic effects do not chance 
the absorbance in the infrared regime.\cite{Yang}  Trevisanutto 
\textit{\textit{et al}.} presented \textit{ab initio} many-body calculations of 
the optical absorption of single and bilayer graphene by solving the 
Bethe-Salpeter equation (BSE) on top of a GW approximation. For graphene, they 
predicted an excitonic resonance at 8.3 eV arising from a background continuum 
of dipole forbidden transitions, and for the bilayer system, the resonance was 
predicted to be shifted to 9.6 eV.\cite{Trevisanutto}

The stacking-dependent optical spectrum of bilayer graphene, particularly 
twisted bilayer graphene, was calculated by Chen \textit{\textit{et al}.} using 
the first-principles GW-BSE approach. They demonstrated that
electron-electron and electron-hole interactions must be included in the 
calculations of the optical spectrum.\cite{Chen} The excitonic 
effects in twisted bilayer graphane with various twist angle was discussed in 
the first-principles GW-BSE scheme by Havener \textit{\textit{et al}.} and they 
showed that including electron-hole interactions was essential to reproduce the 
conductivity spectrum.\cite{Havener} They also found coherent
interactions between states associated with multiple van Hove
singularities in twisted bilayer graphene. Significance of the
stacking order in few-layer graphene was investigated by Yan \textit{et
al.}.\cite{Yan1} Their results suggested that when compared with AB-stacked
few-layer graphene ABC-stacking has significant effects on the optical
absorption spectrum with an absorption peak that appeared around 0.3 eV. As the 
number of layers increases, the absorption amplitude is greatly enhanced, and 
the absorption peak redshifts.

\begin{figure}[htbp]
\includegraphics[width=7cm]{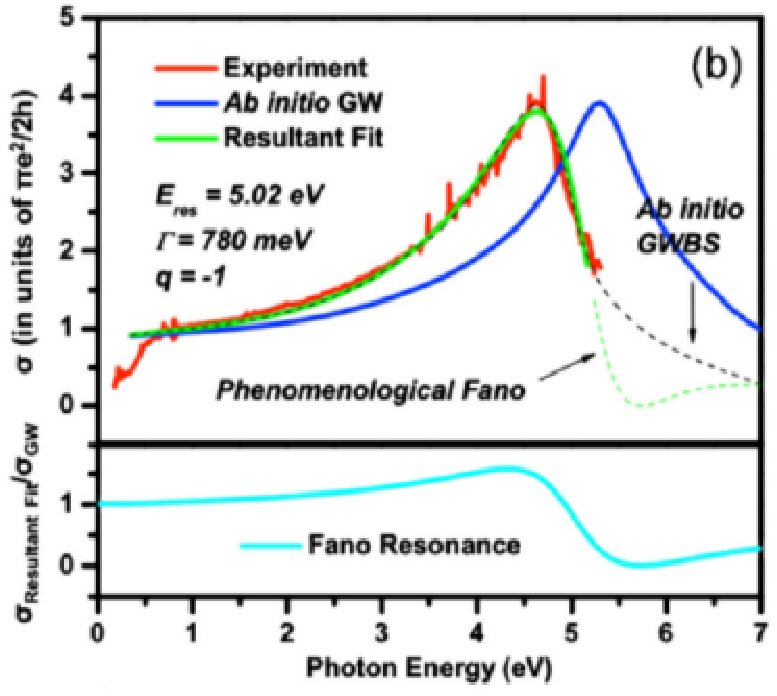}
\caption{\label{mak_fano}
Fit of experiment to the Fano model [from Ref. 70] using the optical 
conductivity obtained from GW calculations\cite{Yang} for $\sigma_{cont}$. The 
dashed line is the optical conductivity spectrum obtained from the full GW-BSE 
calculation.\cite{Yang} }
\end{figure}

Yang \textit{et al}. performed first-principles 
calculations on graphene and bilayer graphene and reported that the 
high-frequency optical spectrum was dominated by both broad and narrow 
resonant excitons. Yang also identified the binding energy of this exciton; 270 
meV in graphene and 80 meV in bilayer graphene.\cite{Yang2} Measurement of Mak 
\textit{et al}. indicated that strong electron-hole interactions 
appeared in the optical conductivity at 4.62 eV, which is redshifted by nearly 
600 meV from the value predicted by \textit{ab initio} GW calculations which is 
shown in Figure \ref{mak_fano}.\cite{Mak} This observation is explained within 
a phenomenological model as a Fano interference effect where the
parameters for that model were extracted from experimental 
measurements.\cite{Fano,Phillips} The Fano model was also used to describe the 
absorption of single- and bilayer graphene by Chae \textit{\textit{et 
al}.}.\cite{Chae} Wachsmuth \textit{\textit{et al}.} performed first-principles 
calculations in the framework of time-dependent DFT (TDDFT) and calculated the 
energy-loss function of single-layer graphene. Despite the disagreement with 
the experiment in the limit of small and vanishing momentum transfers, good 
agreement was found for finite momentum transfers if crystal local-field 
effects were taken into account.\cite{Wachsmuth}

Effects of mechanical and chemical treatments on the optical
properties of graphene were also investigated. First-principles 
calculations were performed to obtain electric field induced band gap and optical 
properties of bilayer graphene and suggested that optical absorbance is found 
to be strongly dependent on the stacking order of bilayer graphene, and the 
polarization direction of the incident light.\cite{Yang3,Mohan}A DFT 
calculation was performed within the \textit{ab initio} spin-density functional 
code AIMPRO\cite{aimpro} by Pereira \textit{et al}. in order to find the optical 
conductivity of graphene under strain.\cite{Pereira} They reported that if the 
system is anisotropically strained, the optical response 
of graphene becomes frequency-independent. Yang \textit{et al}. performed 
first-principles calculations on the optical absorption spectrum of doped 
graphene to reveal many-body effects.\cite{Yang4} These calculations showed 
that charge doping enhances the screening and reduces the e-h interaction, as a 
result, the red shift of the absorbance peak around 4.5 eV is reduced. At 
different doping levels, the energy loss function of graphene at the K Dirac 
cone were calculated within DFT and the changes of low-energy excitations 
around the Dirac cone with the carrier density was explained.\cite{Gao} 
First-principles TDDFT formalism was also used to obtain the excitations near 
the Dirac cone for doped graphene by Despoja \textit{\textit{et al}.} 
\cite{Despoja} and it was reported that their findings agreed with experimental 
results.

Kadi \textit{\textit{et al}.} calculated the optical properties of 
Bernal-stacked bilayer graphene  using a density matrix formalism which 
includes fully momentum-dependent optical and Coulomb matrix elements. A 
noticeable peak is observed in the low energy spectrum of the structure which 
results from the cross transitions at the Dirac point. On the other hand, the 
spectrum in the ultraviolet region showed two close peaks originating from 
interband transitions.\cite{kadi}

Not only doping but also mechanical treatment is capable to tune the optical 
properties. Singh \textit{\textit{et al}.} studied the optical properties of 
clean and oxidized vacancies in graphene by means of DFT 
and indicated that some type of vacancies showed particular peaks in the 
absorption spectrum and it is possible to obtain the oxidation level from optical 
spectroscopy.\cite{Singh} First-principles calculations of the dielectric 
function in the framework of the random-phase approximation (RPA) was carried 
out by Sedelnikova \textit{\textit{et al}.} and showed that out-of-plane bending 
of graphene removes the restriction on the absorption of certain light 
polarization, as a result, the energy position, intensity, and width of the 
peaks of the absorption curves are sensitive to bending.\cite{Sedelnikova}

Graphene nanoribbons (GNR), with different edge formations, are novel 
materials. Because of reduced dimensionality, excitonic effects are dominant in 
the optical spectrum and the binding energy of excitons is also large for 
ribbons which is of importance for optical 
applications.\cite{Yang5,Prezzi,Wang1} Graphene nanoflakes with 
different shapes were also investigated. According to DFT 
calculations, nanoflakes, with van Hove singularities and edge states, 
showed a different optical spectrum in visible light.\cite{Yamamoto,Zhou}  In 
addition to nanoribbons and nanoflakes, the synthesized stable 2D carbon 
allotrope graphdiyne, where the benzene rings are connected by acetylene bonds, 
is dominated by excitonic effects. This was calculated within the GW-BSE and 
was found to be a candidate for optical applications. \cite{Luo,Huang}

\begin{figure}[htbp]
\includegraphics[width=8.5cm]{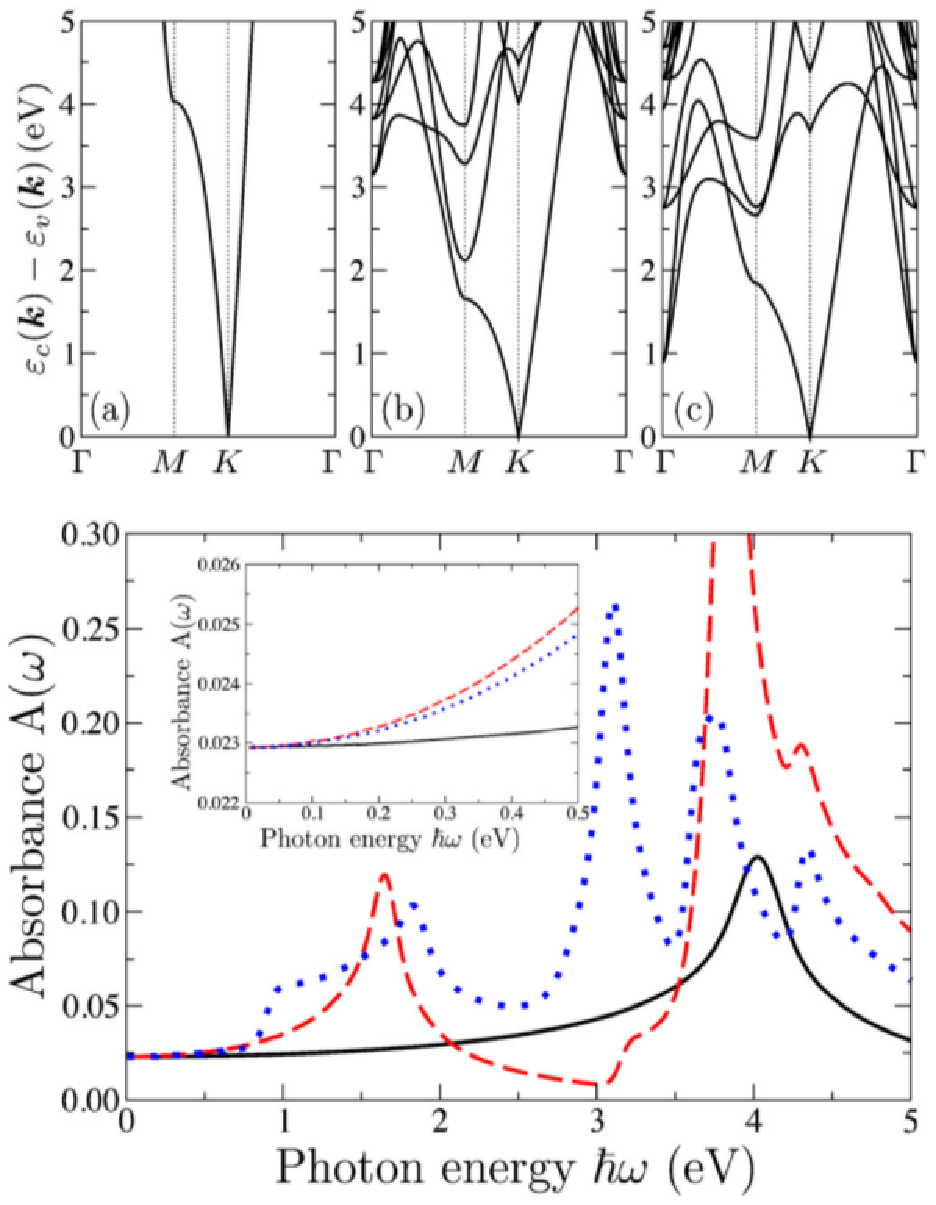}
\caption{\label{silicene_bechstedt_APL_2012}
Interband transition energies along high-symmetry lines in the Brillouin zone 
(BZ) for graphene (a), silicene (b), and germanene (c). Lower panel, 
\textit{ab-initio} calculated optical absorbance of graphene (black 
solid line), silicene (red dashed line), and germanene (blue dotted line) 
\textit{vs} photon energy. The infrared absorbance is shown in the 
inset [from Ref. 92].}
\end{figure}

Silicene and germanene are 2D buckled hexagonal structures constructed from 
group-IV elements (Si and Ge, respectively). Structural and electronic 
similarities with graphene make these materials of interest for their 
optical properties. Bechstedt \textit{\textit{et al}.} studied the optical 
absorbance of graphene, silicene, and germanene by ab-initio calculations 
within the independent-particle approximation. The low energy absorbance was 
found to be the same for all structures as predicted for massless Dirac 
fermions ($A(0)=\pi \alpha$) and showed that the peaks and shoulders in the 
absorption spectrum can almost be related to band edges which are shown in 
Figure \ref{silicene_bechstedt_APL_2012}.\cite{Bechstedt,Matthes}

Cakir \textit{et al.} calculated the optical properties of 
single layer black phosphorus (phosphorene) using GW-BSE method and found that 
the optical response of the monolayer strongly depends on the direction and 
the magnitude of the applied strain. \cite{deniz} It was also demonstrated 
that the stacking of layers and the type of stacking modifies the optical properties 
of multilayer phosphorene. \cite{deniz2} Seixas \textit{et al.} calculated the 
exciton binding energy of phosphorene under strain and found that strain 
modifies the binding energy of the excitons in the monolayer.\cite{seixas} The 
optical response of phosphorene nanoribons was calculated by Tran \textit{et 
al.} and they showed that although armchair phosphorene nanoribons 
have optical transitions close to the band gap, zigzag  phosphorene 
nanoribons have symmetry-gap-forbidden transitions at the band edge. 
\cite{tran}

\subsection{Derivatives of Graphene}

Graphene can be modified by attaching hydrogen or halogen atoms. Such graphene 
derivatives gathered attention over the past years. Graphane is constructed by 
hydrogenation of the C atoms alternatingly above and below the graphene layer. 
Effect of hydrogenation of graphene 
on the optical spectrum was calculated in the independent particle approach of DFT 
by Pulci \textit{\textit{et al}.} They reported that the optical spectrum is
dramatically changed. The electronic band gap of 
graphane is ~3.5 eV while the absorption curve is almost zero up to 7 
eV.\cite{Pulci} Cudazzo \textit{\textit{et al}.} performed first principles 
calculations by solving the GW-BSE for graphane and demonstrated that 
the localized charge-transfer excitations, which were governed by enhanced 
electron correlations, dominated the optical properties of 
graphane.\cite{Cudazzo} Wei \textit{\textit{et al}.} carried out first 
principles calculations within the GW-BSE scheme to obtain the optical 
properties of C$_{4}$H, which is partially hydrogenated graphene and they 
indicated that, as in ordinary graphane, strong charge-transfer excitonic 
effects dominated the optical properties, and the binding energy of the first 
exciton peak was found to be 1.67 eV.\cite{Wei} Cheng \textit{\textit{et 
al}.}\cite{Cheng} focused on half-graphane, where only one sublattice was 
passivated by hydrogen, and investigated also C$_{6}$H to obtain the electronic 
and the optical properties by using the ABINIT code\cite{ABINIT}. They 
reported that the hydrogen atom behaves as an impurity, and this caused an 
enhancement of the optical conductivity.

\begin{figure}[htbp]
\includegraphics[width=7.5cm]{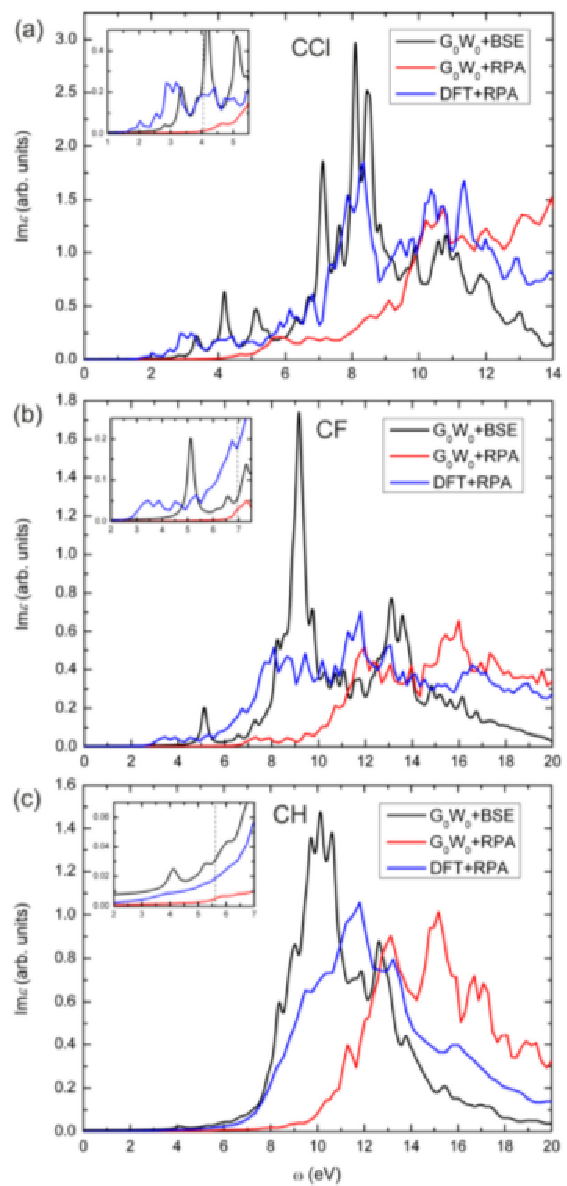}
\caption{\label{derivatives_otyepka_2013}
Absorption spectrum of monolayer (a) CCl, (b) CF, and (c) CH for light
polarization parallel to the surface plane. Insets show amplified regions
of the absorption spectrum in the vicinity of the first exciton peak and
G$_{0}$W$_{0}$ (PBE) gap (dashed line) [from Ref. 103].}
\end{figure}

The optical response of fluorographene was investigated by Samarakoon \textit{\textit{et al}.}  
using the GW-BSE  
approach and they reported that the results were in good agreement with the 
experiments.\cite{Samarakoon} 
Charge-transfer excitations arising  from strong electron-hole interactions were 
obtained. In addition, Karlick\'{y} \textit{\textit{et al}.} calculated 
the optical properties of  chlorographene (CCl), fluorographene (CF), and 
graphane (CH) at the GW-BSE level  and found the formation of exciton  peaks at 
2.82, 5.12, and 4.11 eV with huge binding energies of 1.25, 1.85, and 1.53 eV 
for CCl, CF, and CH,  respectively, as shown in Figure
\ref{derivatives_otyepka_2013}.\cite{karlicky} The importance of excitonic
interaction for single-, double-layer and bulk CF was also studied by
Karlick\'{y} \textit{\textit{et al}.} and they reported additionally 
that the dominant high-energy peak of excitonic absorption systematically shifted to  
higher energies by increasing the number of layers which can be useful for the  
determination of the thickness of samples.\cite{karlicky_2} In addition, Wei 
\textit{et al.} reported that the inclusion of excitonic effects is crucial for 
the determination of the optical properties of CF and C$_{4}$F after performing 
a first-principle  calculation at the level of GW-RPA and GW-BSE.\cite{Wei2} Moreover,
Gunasinghe \textit{et al.}\cite{Gunasinghe} presented DFT calculations 
within the GW-BSE approach as implemented in the YAMBO package.\cite{Yambo} It 
was also indicated that the results of GW-BSE were in good agreement with the 
experiment, and inclusion of the e-h interactions is important for 
description for the excitation effect.

\subsection{Transition-Metal Chalcogenides}

Due to potential applications in the fields of optoelectronics and 
photonics of single layer transition-metal dichalcogenides (TMDs) there has 
been an immense interest in the investigation of their optical properties. 
Since they are expected to have strong excitonic effects due to the weak 
dielectric screening which will strongly influence their optical 
properties, it is vital to include these effects into the calculations.

MoS$_2$ is one of the most investigated 2D materials among the TMDs. It is an 
indirect gap semiconductor in its bulk form with a band gap of around 1 eV 
\cite{gourmelon} and  it becomes a direct gap semiconductor
when it is isolated as a monolayer.  \cite{mak,splendiani} Monolayer TMDs and 
particularly MoS$_2$ is a challenging system for calculating the optical 
properties because the results are quite sensitive to the settings such as the 
number of empty bands (N$_{b}$),  the energy 
cutoff for the response function (E$_{r}$) and the sampling of the Brillouin 
zone used in the calculations. That is why different settings
lead to different results for the optical spectrum. It has been shown that 
the dependence of the optical spectrum on the calculation settings are more 
prominent when the quasiparticle wave functions contain localized states and 
the conduction band minimum (CBM) and valence band maximum (VBM) have different 
orbital character.\cite{shih} It is known that the VBM of monolayer MoS$_2$ at K 
and K$^{'}$ is strongly delocalized by the $d_{x^{2}-y^{2}}$ orbital from the 
Mo atom, and the $p_{x}$ and $p_{y}$ orbitals from the S atom. On the other 
hand, the CBM has $d_{z^{2}}$ character from the Mo atom and a small 
contribution from $p_{x}$ and $p_{y}$ orbitals of the S atom. \cite{cao} This 
makes the system computationally challenging.

\begin{table*}
\caption{\label{table1} The positions of A, B, A$^{'}$, B$^{'}$, and C exciton 
peaks from experiments and several theoretical calculations. The convergence 
criterion i.e. used k mesh, number of valence (N$_v$) and conduction bands 
(N$_c$), the energy cutoff for the response function (E$_r$) and the number of 
empty bands (N$_b$) used in the calculations are also listed. }
\begin{tabular}{l|cccccccccccccc|cccccccccccc}
\hline\hline
                            &     \multicolumn{14}{c}{Optical Transition (eV)}  
                     & & & \multicolumn{5}{c}{Convergence Parameters}               
                                                       \\
\hline
                            &    &   A  &    &   B  &     &   A$^{'}$ &    &  
B$^{'}$ &    &   C   & &   E$_b$  & &E$_g$GW& & k-mesh            & &    N$_v$  
&         & N$_c$   &   & E$_r$(eV) &  &N$_b$        \\
Absorp. Exp.\cite{mak}      &    & 1.88 &    & 2.03 &     &   --      &    &   
--     &    &  --   & &    --    & & --      & &  --               &  &     --

&         &  --     &   &  --       &  & --          \\
PL Exp.\cite{splendiani}    &    & 1.85 &    & 1.98 &     &   --      &    &   
--     &    &  --   & &    --    & & --      & &  --               & &     --   
 
&         &  --     &   &  --       &  & --          \\
G$_0$W$_0$-BSE\cite{ashwin} &    & 1.78 &    & 1.96 &     &   --      &    &   
--     &    &  --   & &    1.04  & &  2.82 & &$6\times6\times1$  & &     2     
&         &  4      &   &  272      &  & 96          \\
sc-GW$_0$-BSE\cite{shi}     &    & 2.22 &    & 2.22 &     &   2.50    &    &   
2.50   &    &  --   & &    0.63  & &  2.80 & &$15\times15\times1$& &     6     
&         &  8      &   &  300      &  & 197         \\
G$_1$W$_0$-BSE\cite{qui}    &    & 1.88 &    & 2.02 &     &   2.20    &    &   
2.32   &    & 2.54  & &    0.96  & &  2.84 & &$72\times72\times1$& &     7     
&         &  8      &   &  476      &  & 6000        \\
\hline
\end{tabular}
\end{table*}

It is known from experiments that the positions of the A and B exciton peaks in 
MoS$_2$ monolayer are around 1.85 eV and 1.98 eV, respectively. 
Band structure calculations show that this splitting is due to the 
effect of spin-orbital coupling (SOC) at the K point of the BZ. It was also found
in the literature that the photoluminescence peak increases from bulk 
to monolayer MoS$_2$, which is consistent with the indirect-to-direct 
band gap  crossover as the sample is reduced to a monolayer.  

\begin{figure}[htbp]
\includegraphics[width=7cm]{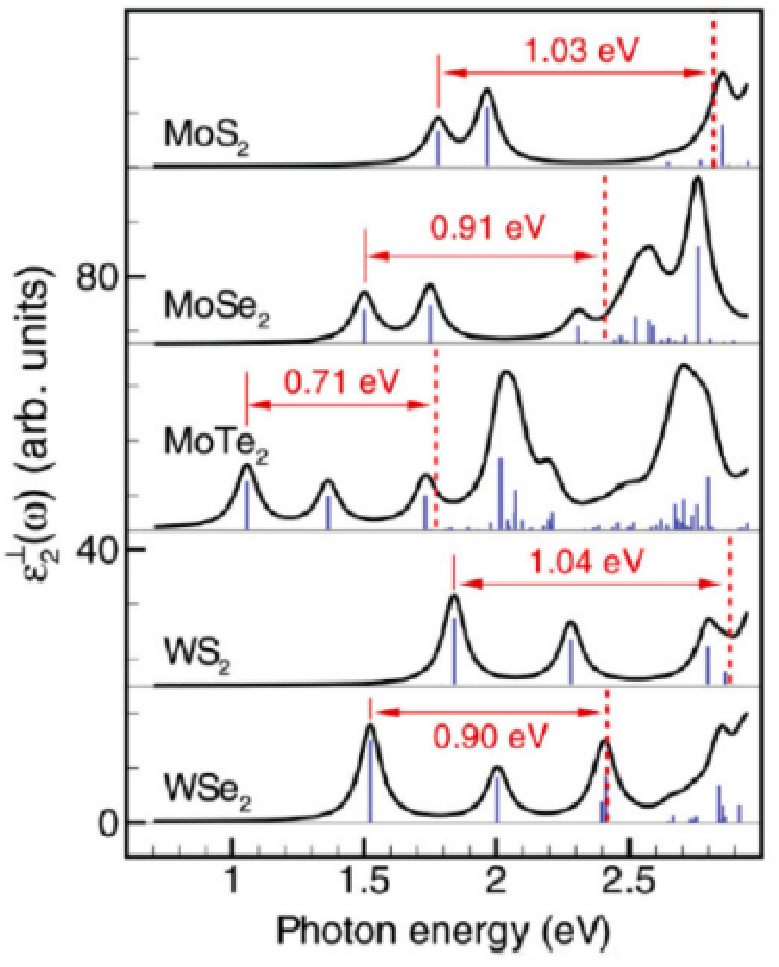}
\caption{\label{ET3}Imaginary part of the  dielectric constant for monolayer (a)
MoS$_2$,  (b) MoSe$_2$, (c) MoTe$_2$, (d) WS$_2$,  and  WSe$_2$. The blue lines 
correspond to  the relative oscillator strengths for the optical transitions. 
The dashed red lines indicate the G$_0$W$_0$ band gap. The binding energy of A 
exciton is indicated in the figure [from Ref. 111]).}
\end{figure}

\begin{figure}
\includegraphics[width=8.5cm]{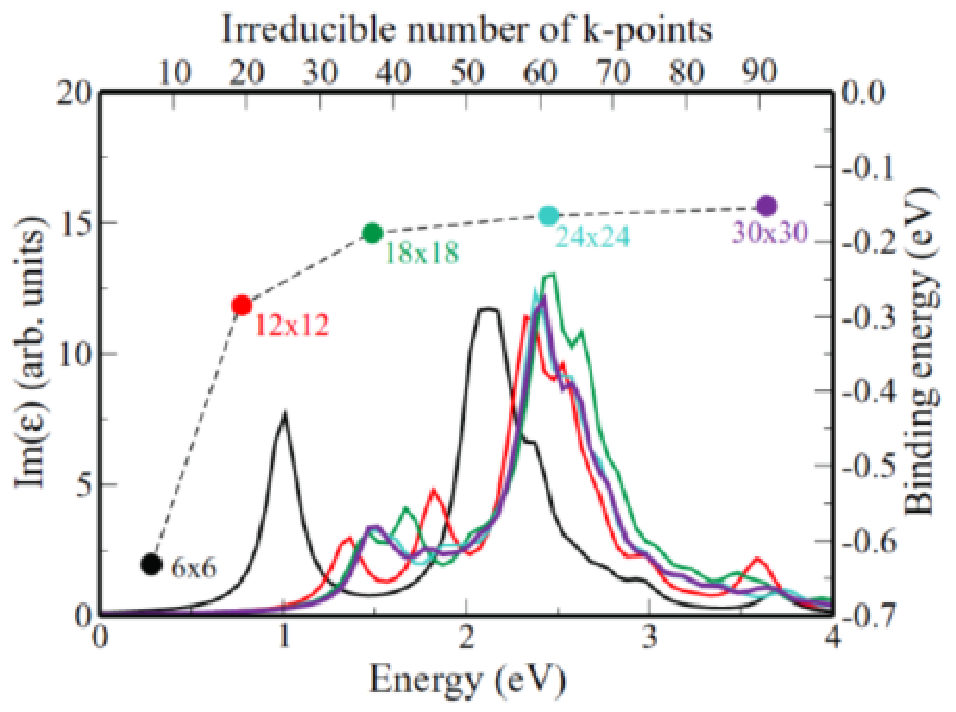}
\caption{\label{ET2} Imaginary part of the dielectric function, obtained by 
solving BSE, of MoS$_2$ for several k-point meshes (Left axis). Binding energy 
of the exciton (E$_{b}$) as a  function  of  the  number  of  irreducible 
k-points (Right axis) [from Ref. 113].}
\end{figure}

In the literature there are several works on the absorption spectrum 
of  monolayer MoS$_2$. Cheiwchanchamnangij \textit{\textit{et al}.} showed that 
by using the self consistent GW method with Mott-Wannier theory, the MoS$_2$ is a 
direct gap semiconductor.\cite{cheiwchanc} The calculation was performed with a 
k-mesh of $8\times8\times2$ for monolayer MoS$_2$ and it was found that 
the transition energy values for A and B excitons are 2.75 and 2.9 eV which is 
slightly overestimated  as compared to the experimental values. Similarly, 
Ramasubramaniam showed that the MoS$_2$ is also an direct gap semiconductor by 
using the single-shot GW method (G$_0$W$_0$) on top of Heyd-Scuseria-Ernzerhof
(HSE) hybrid calculations which is shown in Figure~\ref{ET3}.\cite{ashwin} The 
calculation was performed with a 
$6\times6\times1$ k-mesh, 272 eV for the E$_r$ and the N$_b$ value was taken as 
96. 2 highest valence bands (N$_v$) and 4 lowest conduction bands (N$_c$ ) were 
included as basis for the excitonic state at the BSE level. With these settings 
the positions of A and B excitons were found to be quite similar to the
experimental values, 1.78 and 1.96 eV, respectively and with an exciton binding 
energy of 1.04 eV(see Table \ref{table1}). By using self-consistent GW method, 
Shi \textit{\textit{et al}.} showed that the monolayer MoS$_2$ is an indirect 
band gap semiconductor at the G$_0$W$_0$ level and updating the G part makes it 
a direct band gap semiconductor.\cite{shi} In this calculations a 
$15\times15\times1$ k-mesh is used together with 300 eV for E$_r$, N$_b$ value 
is taken as 197 and the N$_v$ and N$_c$ are selected as 6 and 8, respectively. 
In this calculation in addition to A and B exciton peaks (at 2.22 eV), the 
A$^{'}$ and B$^{'}$ exciton peaks (at 2.50 eV) are also found which are not 
experimentally active (see Table \ref{table1}). The exciton binding energy was 
calculated as 0.63 eV in this study. Noticed that results exhibit clear 
discrepancies due to the different settings used in the calculations. For 
instance the dependence of the optical spectrum of the monolayer MoS$_2$ on the
Brillouin zone sampling is shown in Figure \ref{ET2}. As can be seen the 
position of the peaks are moving towards higher energy values with increasing 
the k-point grid.\cite{sangalli,shi}

\begin{figure}[htbp]
\includegraphics[width=8.5cm]{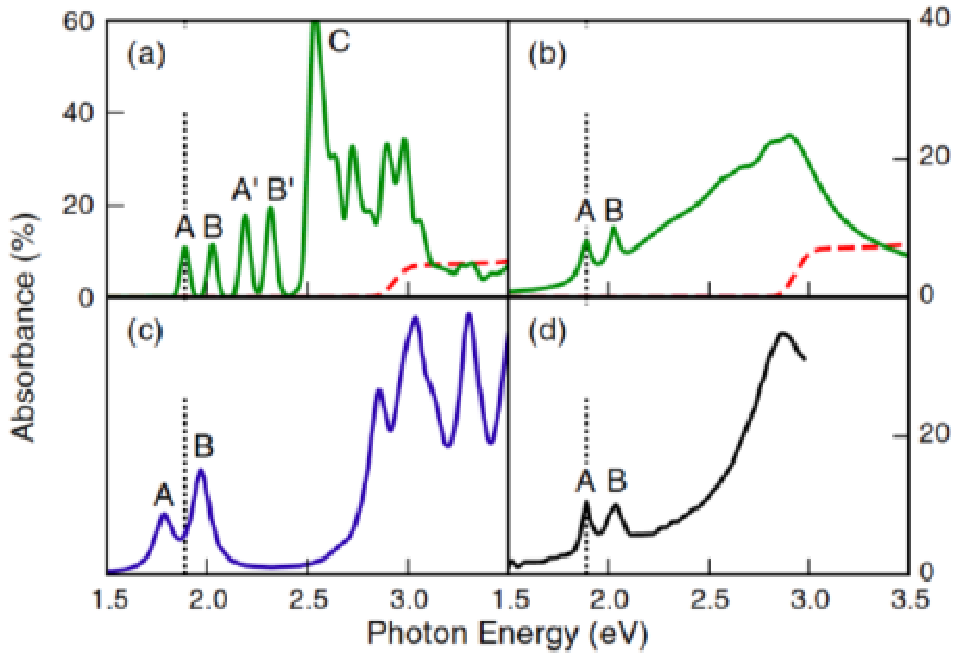}
\caption{\label{ET4}(a) Absorption spectrum of MoS$_2$ without (dashed red 
curve) and with (solid green curve) electron-hole interaction 
using a constant broadening of 20 meV. (b) Same data using an \textit{ab 
initio} broadening based on the electron-phonon interaction. (c) Previous 
calculation using G$_0$W$_0$ (d) Experimental data [from Ref. 114].}
\end{figure}

In order to address these discrepancies, a more recent work by Qiu \textit{et 
al.}  calculated the absorption spectrum of monolayer MoS$_2$ with very high 
settings.\cite{qui} In this study the G  part is updated self-consistently one time 
(G$_1$W$_0$) because an additional iteretion change the
the band gap by less then 20 meV. A $72\times72\times1$ k-mesh is 
used with energy cutoff of E$_{r} =$ 476 eV and N$_{b} =$ 6000 bands where both 
parameters must be converged together for the accurate QP energies. The number 
of valance bands N$_v$ and conduction bands N$_c$ are selected as 7 and 8 for 
BSE calculation, respectively. It was predicted that there are 
exciton peaks in the 2.2 and 2.8 eV range. These peaks were not found in other 
theoretical and experimental works. Calculated absorption spectrum is shown in
Figure \ref{ET4}. They suggested that the previous calculations 
\cite{ashwin,shi} were not converged with respect to k-point sampling, and the 
agreement between the peak position for A and B and experimental results
found in earlier reports are likely accidental. It was also suggested by Qiu \textit{et al.} that 
the inconsitancy between 
the converged calculations and the experiment can be solved by considering the quasiparticle 
lifetime effect due to 
electron-phonon interaction which is shown in Figure \ref{ET4} (b). 

In addition to MoS$_2$, Ramasubramaniam \textit{\textit{et al}.} calculated 
the optical properties of MoSe$_2$, MoTe$_2$, WS$_2$, WSe$_2$ and WTe$_2$ 
solving BSE on top of G$_0$W$_0$ and HSE hybrid functional  method. It was 
reported that these materials have 
strongly bound excitons which have binding energies around 1 eV. These binding 
energies are becoming smaller and exciton peaks show a systematic
redshift as the chalcogen atoms get heavier.\cite{ashwin} The absorption 
spectrum together with the oscillator strength (shown by  blue lines) are shown 
in Figure \ref{ET3}. The height of the lines are a measure of the oscillator 
strength.

The binding energy of excitons in monolayer and  bilayer MoS$_2$ was also 
calculated by using Mott-Wannier theory and reported as 0.897 eV and 0.424 eV, 
respectively. \cite{cheiwchanc} As seen the exciton binding energy is smaller 
in bilayer MoS$_2$.

The strain dependent optical absorption spectrum of single layer MoS$_2$ was  
investigated by using BSE calculations on top of G$_{0}$W$_{0}$ \cite{feng} 
and self-consistent GW$_{0}$ \cite{shi} calculations. The reported binding 
energies of excitons are smaller than the previous values which are around 0.5 
eV and it was reported that these binding energies were not affected by biaxial 
strain.

The optical properties of other nanostructure forms of TMDs, such as MoS$_2$ 
nanoflakes, were also investigated by using the density 
functional tight-binding (DFTB) method. It has been shown that with increasing 
edge length of the structures the absorption peaks are red-shifted which is due 
to the quantum confinement effect. \cite{joswig} Jing \textit{et al.} showed 
that adsorbing organic molecules on MoS$_2$ leads to a considerable charge 
transfer from the adsorbed molecule to the MoS$_2$ layer. This transfer changes 
the electronic structure of the material but not much effect on the absorption 
spectrum of the system.\cite{jing}

The optical properties of low dimensional materials have are strongly correlated 
with the surrounding dielectric environment. In monolayer MoS$_2$, this 
correlation leads to a blue shift up to 40 meV in the photoluminescence (PL) 
peaks.  Lin \textit{et al}.  analyzed this phenomena using an analytical 
investigation and showed that the surrounding dielectric environment has an 
effect on the exciton and trion binding energies of monolayer MoS$_2$.\cite{lin}

Vacancies and defects in TMD monolayers changes the optical properties. Wei 
\textit{et al.} showed that atomic defects in WS$_2$ monolayers lead to an 
increase in the absorption intensity and a red shift in the absorption spectrum 
of the material by using first-principle methods.\cite{wei} An analytical 
investigation showed the effect of point defects in MoS$_2$ and WS$_2$ 
monolayers on the optical 
properties of the materials. It was shown that the localized charge 
carriers around these point defects modify the optical characteristics of the 
systems.\cite{yuan} Feng \textit{et al.} showed that the static dielectric 
constant of monolayer MoS$_2$ decrease with Mo vacancies and increase with S 
vacancies by using first-principle calculations.\cite{pingfeng} The stacking of 
these TMDs has also an influence on the optical spectrum of these materials. He 
\textit{et al.} investigated this phenomena by using BSE calculations on top of 
G$_{0}$W$_{0}$ with spin-orbit coupling for bilayer  MoS$_2$, MoSe$_2$, 
WS$_2$, and WSe$_2$. They showed that stacking has an effect on the exciton 
binding energies of these systems.\cite{he}

\begin{figure*}[htbp]
\includegraphics[width=12cm]{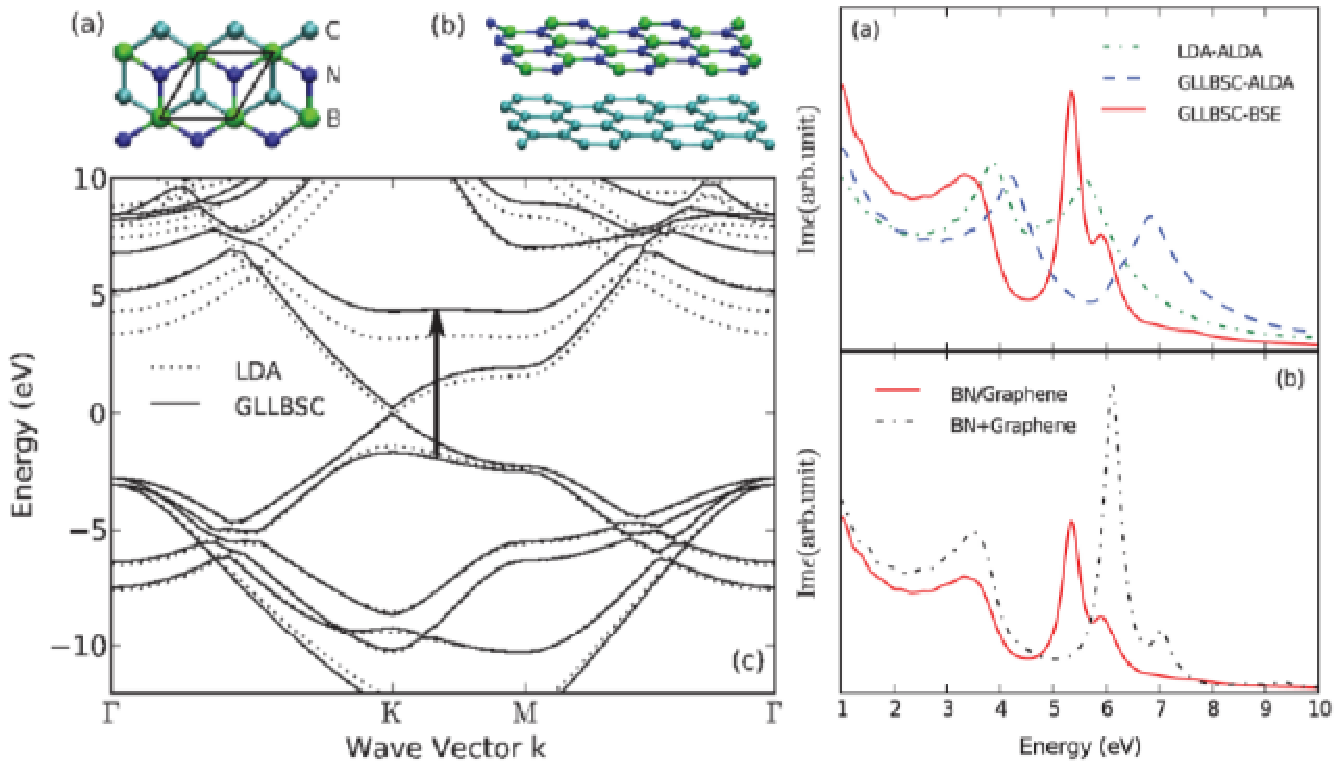}
\caption{\label{graphene-bn_yan_prb_2012}
(a) Top and (b) side view of a graphene/h-
BN. (c) Band structure of graphene/h-BN calculated with GLLBSC
(solid lines) and LDA (dotted lines). The top of the valence bands is
set to zero. Upper right panel: Optical absorption spectrum
of graphene/h-BN calculated using the LDA-ALDA (dash-dotted
line), GLLBSC-ALDA (dashed line), and GLLBSC-BSE (solid line).
Lower right panel: The GLLBSC-BSE spectrum of the interface (repeated)
together with the sum of the absorption spectrum of an isolated graphene
and h-BN layer, respectively [from Ref. 122].}
\end{figure*}

\subsection{Heterostructures made of Monolayer Building Blocks}

The optical properties of heterostructures which are made of MX$_2$, graphene 
and hexagonal boron nitride monolayers are also of interest due to their 
unique optical properties. Even preliminary results on such materials as 
mentioned below have revealed the possibility of using heterostructures in 
various optoelectronics device applications.  

The counterpart of graphene, \textit{h}-BN is a well-known 2D crystal that 
differs from graphene with its insulating properties. Heterostructures of both 
materials are promising for not only electronic but also for advanced optical 
applications. Calculations of Yan \textit{\textit{et al}.} by including e-h 
interactions through the BSE-RPA showed that the absorption spectrum of the 
\textit{h}-BN-on-graphene system is not simply a sum of the absorptions of the 
isolated layers which is shown in Figure \ref{graphene-bn_yan_prb_2012}, 
because the transition energies in \textit{h}-BN become redshifted by $\sim1$ 
eV due to the screening by the graphene electrons.\cite{Yan} Attaccalite 
\textit{\textit{et al}.} focused on the effect of defects on the optical 
properties and reported that ``deep level impurity'' peaks in the optical 
spectrum are strongly affected by the e-e and e-h interaction and can be 
reliably calculated only on the level of many-body perturbation 
theory.\cite{Attaccalite}  Wang \textit{\textit{et al}.} investigated the 
possible hyperlens application of graphene and \textit{h}-BN by using
DFT.\cite{Wang} They reported that both graphene and \textit{h}-BN exhibit
strong anisotropic properties in the ultraviolet region, where their 
permittivity components perpendicular to the optical axis can be negative, 
while the components parallel to the optical axis can be positive, which is 
desirable for hyperlens implementations.

The optoelectronical properties of lateral hybridized graphene and hexagonal 
boron nitride was also investigated by solving BSE on top of GW calculations. 
\cite{bernardi} It is shown that the exciton binding energy decreases with 
increasing the carbon domain size in the system. The vertical heterostructure 
made of graphene and boron nitride (BN) was investigated by solving BSE and 
using functional by Gritsenko, Leeuwen, Lenthe, and Baerends (GLLBSC).  It is 
predicted that the absorption spectrum of heterostructures is not equal to the 
sum of the individual absorption spectra.

However, heterostructures of TMDs provide another rich playground. Komsa 
\textit{\textit{et al}.} calculated the optical 
absorption spectrum of vertical MoS$_2$/WS$_2$ heterostructure by solving BSE 
equation on  top of single shot G$_0$W$_0$ calculation \cite{komsahet} and it 
was observed that the optical spectrum of such a heterostructure has the same 
features as the individual monolayers and the optically active transitions are 
formed by direct intralayer transitions.  

Recently Kang \textit{et al}. showed that certain van der Waals 
heterostructures, can facilitate spatial separate electron-hole localization which 
may find applications in photovoltaic energy conversion. \cite{Kang} The 
electronic structure of various heterostructures were calculated in the 
GW level by Leb\'{e}gue  \textit{et al}. They predicted that due to 
the band alignment, the MX$_2$ heterostructures can be used as 
charge separator for solar cell applications. \cite{debbichi}  The exciton 
radiative lifetimes of MoS$_2$/WS$_2$ and MoSe$_2$/WSe$_2$ bilayer 
heterostructures are calculated by solving the BSE equation  on top of GW 
calculation. It is predicted that these heterobilayers exhibit long-lived ($\sim 
20-30$ ns at room temperature) excitons whose electron is localized on the Mo and 
the holes are localized on the W based layer. \cite{palummo}

\begin{figure*}[htbp]
\includegraphics[width=7 cm]{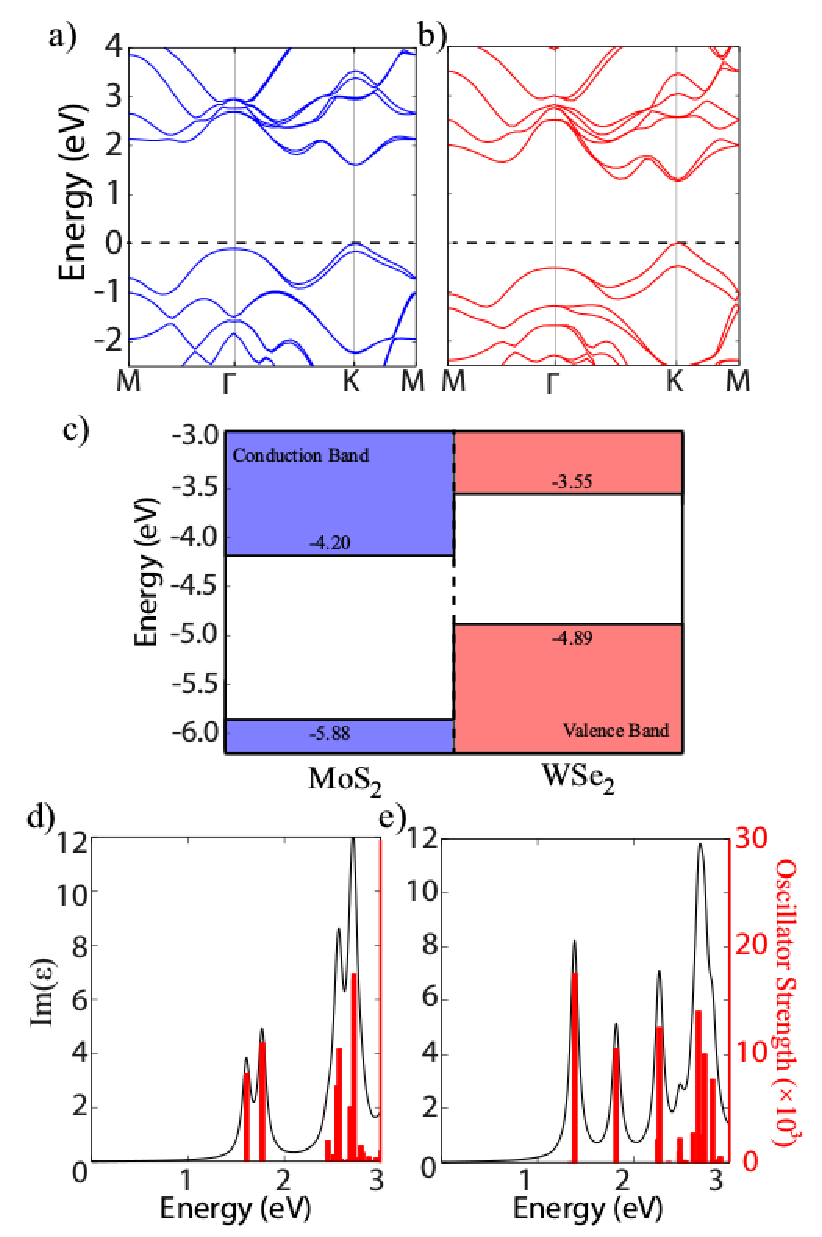}
\caption{\label{fig13}The band structure of (a) MoS$_2$  and (b) WSe$_2$ with 
SOC, Fermi level is set to 0 eV. (c) Calculated band aligment for  MoS$_2$  and 
WSe$_2$ monolayers. Imaginary part of the 
dielectric function for (d) MoS$_2$ and (e) WSe$_2$ together with oscillator 
strenght of the optical transitions. }
\end{figure*}

Recent experiments on vdW heterostructures of TMDs have also 
revealed fingerprints of some unusual properties in their photoluminescence 
spectrum. In Figure \ref{fig13}, we present our first-principle calculations on 
MoS$_2$ and WSe$_2$ monolayers. As seen from the band structures both 
monolayers are direct band gap semiconductors. The band gap of the MoS$_2$ 
monolayer is slightly larger than the band gap of the 
WSe$_2$ monolayer and the splitting at the K point in the Brillouin zone is 
more apparent for WSe$_2$ due to having a larger SOC. The band alignment 
calculations show that a possible heterostructure formed from these monolayers 
would be a type II heterojunction. This leads to spatially indirect excitons 
due to localization of the electrons and holes in different monolayers of the 
heterostructure. The recombination of the electron hole pair occurs through the 
staggered gap of the heterojunction. In Figure \ref{fig13}(d) and (e), the 
imaginary part of the dielectric function and oscillator strength of the 
optical transitions of MoS$_2$ and WSe$_2$ monolayers are shown, respectively. 
To obtain the dielectric function of these monolayers, BSE calculation on top of 
single shot GW calculation (G$_0$W$_0$) calculation is performed. 
Our calculations revealed that the oscillator strengths of the optical 
transition for MoS$_2$ monolayer is smaller than the oscillator strengths of 
the WSe$_2$ monolayer which fits with the earlier experimental findings 
performed by Fang \textit{et al.} \cite{battaglia} In the same work it has been 
also shown that the oscillator strength of the optical transitions for 
MoS$_2$-WSe$_2$ heterostructure resembles the oscillator strength of the MoS$_2$ 
monolayer. This shows that the oscillator strength of the heterojunction is 
dictated by the constituent monolayers which has lower oscillator strength.

Despite the increasing number of experimental studies on the optical properties 
of heterostructures, there are still few theoretical works due to the 
computational expense of such calculations. 

\section{Conclusions}
In conclusion, advances in theoretical calculations on the optical properties of atomically thin 
layered crystal structures 
are reviewed. 2D crystals have been a topic of interest for more than a decade due to their unique 
properties which differ 
from their 3D counterparts. However, still significant research efforts are needed for deeper 
understanding of their unique 
optical properties. Regarding the difficulties in synthesis and characterization of materials at 
nanoscale, theoretical 
calculations that provides cheap, fast and reliable tool for analysis of ultra-thin crystal 
structures are of crucial 
importance. This overview reveals the power of theoretical predictions on the optical properties of 
graphenelike ultrathin 
crystal structures.

\section{acknowledgments}
This work was supported by the Flemish Science Foundation (FWO-Vl) and the 
Methusalem foundation of the Flemish government. Computational resources were 
provided by TUBITAK ULAKBIM, High Performance and Grid Computing Center 
(TR-Grid e-Infrastructure). H.S. is supported by a FWO Pegasus Long Marie Curie 
Fellowship. J.K. is supported by a FWO Pegasus short Marie Curie Fellowship.

\end{document}